\documentclass[12pt,a4paper]{article}
\usepackage[english]{babel}
\usepackage[T1]{fontenc}
\usepackage{graphicx}
\usepackage{amsmath, amsthm, amssymb, mathrsfs}
\usepackage{epstopdf}
\usepackage{caption}
\usepackage{subcaption}
\usepackage{float}
\usepackage{appendix}
\usepackage{tikz}
\usetikzlibrary{shapes,positioning}
\usepackage{braket}
\usepackage{anyfontsize}
\usepackage{mathptmx}
\usepackage[11pt]{moresize}
\usepackage{enumerate}
\usepackage{dcolumn}
\usepackage{bm}
\usepackage{tabularx}
\usepackage{array}
\usepackage{lipsum}
\usepackage{cancel}
\usepackage{enumitem}
\usepackage{soul}
\usepackage{authblk}
\usepackage{pgfplots}
\usepackage{cite}

\def\cN{{\mathcal N}}

\newcommand{\bb}{\boldsymbol}
\newcommand{\be}{\begin{equation}}
\newcommand{\ee}{\end{equation}}

\newcommand{\ba}{\begin{align}}
\newcommand{\ea}{\end{align}}

\newcommand{\bea}{\begin{eqnarray}}
\newcommand{\eea}{\end{eqnarray}}
\newcommand{\bt}{\begin{theorem}}
\newcommand{\et}{\end{theorem}}
\newcommand{\bp}{\begin{proposition}}
\newcommand{\ep}{\end{proposition}}
\newcommand{\bd}{\begin{definition}}
\newcommand{\ed}{\end{definition}}

\newtheorem{theorem}{Theorem}
\newtheorem{proposition}{Proposition}
\newtheorem{definition}{Definition}

\begin{document}

\title{Quantum reading of quantum information}

\author[1]{Samad Khabbazi-Oskouei}
\author[2,3]{Stefano Mancini}
\author[2,3]{Milajiguli Rexiti}

\affil[1]{Department of Mathematics, Varamin-Pishva Branch Islamic Azad University, Varamin, 33817-7489 Iran}
\affil[2]{School of Science and Technology, University of Camerino, Via Madonna delle Carceri 9, I-62032 Camerino, Italy}
\affil[3]{INFN–Sezione di Perugia, Via A. Pascoli, I-06123 Perugia, Italy}

\maketitle	

\begin{abstract}
We extend the notion of quantum reading to the case where the information to be retrieved, which is encoded into a set of quantum channels, is of quantum nature. We use two-qubit unitaries describing the system-environment interaction, with the initial environment state determining the system’s input-output channel and hence the encoded information.
The performance of the most relevant two-qubit unitaries is determined with two different approaches: 
i) one-shot quantum capacity of the channel arising between environment and system’s output;
ii) estimation of parameters characterizing the initial quantum state of the environment.  
The obtained results are mostly in (qualitative) agreement, with some distinguishing features that include the CNOT unitary.
\end{abstract}

%\date{\today}

\section{Introduction}

Quantum reading is the process of retrieving classical information from a memory by using a quantum probe 
\cite{P11} (for a survey on the subject see \cite{QR}).
It is customary to see such information encoded into a finite set of quantum channels, each one labeled by the value that a random discrete variable can take. As a such, the process involves quantum channel discrimination \cite{GLN,Sacchi,WY,Hay}.
Quantum reading has been applied in various contexts, ranging from physical imaging \cite{Qim}, to radar \cite{QRA}, to biology \cite{gehring}, to cryptography \cite{Qcryp}, and showed advantages over classical reading.

A prototypical model for a memory cell in quantum reading is the environment parametrized quantum channel \cite{SSA,DW19,RM17,KMWa,KMWb}. This is a unitary acting on two systems: the main system and the environment. Depending on the state of the latter, a channel will be realized on the former by tracing out the environment at the end. Thus, the initial state of the environment can be considered as the encoded information that has to be retrieved (while the main system plays the role of the probe). This model was employed with environment input states forming an orthonormal basis for the associated Hilbert space. In such a way it realized an incoherent model of memory cell (investigated also for private reading \cite{BDW}), as well as a coherent model of memory cell (allowing entanglement generation between encoder and reader \cite{DBW}).

Here, we go beyond the assumption made on the initial quantum state of the environment and consider a generic one to be determined.
Thus, as figure of merit it is natural to resort to the one-shot quantum capacity \cite{BD10} of the channel connecting the environment with the output of the main system (probe). In fact quantum information has to be extracted with a finite number of usage of the channel realized on the main system (probe) and a nonzero probability of error. Since explicit computation of one-shot capacity is challenging, we shall compute a lower bound on it following \cite{PRM18}.

Additionally, we shall consider the information to be retrieved as residing 
on the parameters characterizing the encoded quantum state, thus tracing back the problem to continuous
multi-parameter estimation.
In this case as figure of merit we shall consider a Bayesian version of the quantum Cramer-Rao bound
derived from the classical bound \cite{GL95}. 

We shall confine our attention to two-qubit unitaries. In particular those that are entangling. These can be represented by points lying in a tetrahedron in $\mathbb{R}^3$ (see e.g. \cite{MrSm}). Specifically, we characterize the quantum reading of quantum information on the edges of this tetrahedron starting from its vertices. The found results show a large qualitative agreement between the approach based on the one-shot quantum capacity bound and the approach based on Bayesian quantum Cramer-Rao bound, with some distinguishing features that include the CNOT unitary.

%%%%%%%%%%%%%%%%%%%%%%%%%%%%%%%%%%%%%%%%%%%%%%%%%%%%%%%%%%%%%

\section{The model}

\begin{figure}[H]
\begin{center}
\begin{tikzpicture}[scale=0.5]
\draw[thick] (-2.5,1.5) -- (-0.5,3.0);
\draw[thick] (-2.5,1.5) -- (-0.5,0);
\draw[thick] (-0.5,3) -- (2,3);
\draw[above](0.5,3) node{$R$};
\draw[thick] (-0.5,0) -- (2,0);\draw(-3.3,1.5) node{$|\psi_\phi\rangle$};\draw[above](0.5,0) node{$A$};
\draw[above](6.5,0) node{$B$};\draw(-0.5,-0.5) node{$\phi$};
\draw[thick](2,1)rectangle(5,-3); \draw(3.5,-1) node[font = \fontsize{15}{15}]{$U$};
\draw[thick] (-0.5,-2) -- (2,-2);\draw(-1,-2) node{};\draw[above](0.5,-2) node{$E$};
\draw[dashed] (5,-2) -- (7,-2);
\draw[above](6.5,-2) node{$F$};
\draw[thick] (5,0) -- (7,0);\draw(7,0) node[right][font = \fontsize{15}{15}\sffamily\bfseries]{};
\end{tikzpicture}
\end{center}
 \caption{General model for quantum reading based on environment parametrization of quantum channels. 
 $|\psi_\phi\rangle$ is the purification of the input $\phi$ to the $A$ system.}
      \label{fig1}
\end{figure}
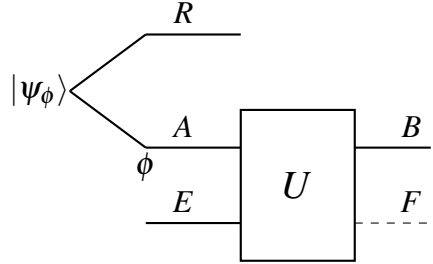

Consider a unitary $U^{AE\to BF}$ with systems $A\simeq B$ and $E\simeq F$. 
By referring to Fig.\ref{fig1}, the environment parametrization of quantum channel \cite{SSA,DW19,RM17,KMWa,KMWb} consists in characterizing a channel ${\cal N}^{A\to B}_\theta$ between $A$ and $B$ in terms of the $E$ state $\theta$.

For the purpose of quantum reading systems $A$ and $B$, i.e. input and output of ${\cal N}^{A\to B}_\theta$, are both held by the reader, which wants to retrieve the state $\theta$ of the system $E$.
It is customary to consider $x\in{\cal X}$ (discrete and finite alphabet) encoded by $E$ as orthogonal states $|x\rangle$ with probability $p_x$. Then, the objective for the reader, given ${\cal N}_x^{A\to B}$, is to find $x$ among all possible values in $\cal X$. This task can give rise to classical communication \cite{Petal11} as well as to quantum communication \cite{DBW} between $E$ and $B$.

In fact, for classical information transmission, an incoherent picture is used leading to the final state of the whole system as 
\begin{equation}
U \left( |\psi_\phi\rangle\langle\psi_\phi | \otimes \sum_x p_x |x\rangle\langle x|\right) U^\dag.
\end{equation}

Instead, for quantum information transmission, a coherent picture is used leading to the final state of the whole system as 
\begin{equation}
\sum_x \sqrt{p_x} |x\rangle V^x |\psi_\phi\rangle,
\end{equation}
where $V^x\equiv U|x\rangle$ is an isometry from $A$ to $BF$.\footnote{This picture was then extended to coherent quantum channel discrimination \cite{MMW20}.}

Here we change the paradigm and according to Fig.\ref{fig1} we consider a generic state $\theta$ encoded into $E$. Then the objective is to recover such a state. This of course implies the necessity of transmitting quantum information from $E$ to $B$ or alternatively the necessity of estimating the environment state.

\bigskip

We will focus on two-qubit unitaries $U^{AE\to BF}$, with states $\boldsymbol{\phi}$, $\boldsymbol{\theta}$ 
on $A$ and $E$ systems respectively. The former is the input of the probe system and the latter the encoded state in the memory. The bold symbols emphasize that the qubit states are characterized by vectors of $\mathbb{R}^3$.

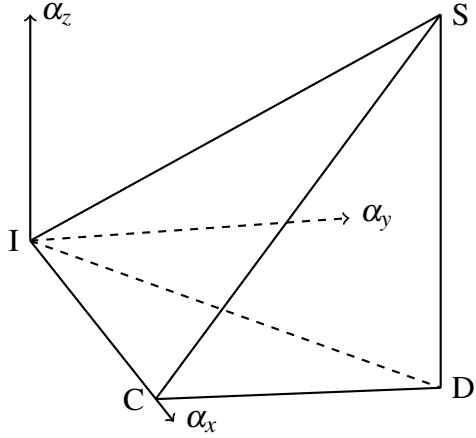
\begin{figure}[H]
\begin{center}
\begin{tikzpicture}[scale=0.3]
\draw[thick,dashed][->] (0,0) -- (14,1);
\node[right] at (14,1){$\alpha_y$};
\draw[thick][->] (0,0) -- (0,10);
\node[right] at (0,10){$\alpha_z$};
\draw[thick][->] (0,0) -- (6.3,-8);
\node[right] at (6.3,-8){$\alpha_x$};
\draw[thick,dashed] (0,0) -- (18,-6.5);
\draw[thick] (5.5,-7) -- (18,-6.5);
\draw[thick] (18,10) -- (18,-6.5);
\draw[thick] (18,10) -- (5.5,-7);
\draw[thick] (18,10) -- (0,0);
\node[left] at(0,0) {I};
\node[right] at(18,10) {S };
\node[right] at(18,-6.5) {D};
\node[left] at(5.5,-7) {C};
\end{tikzpicture}
\end{center}
\caption{Tetrahedron representing the parameters space of two-qubit unitaries, where I, C, S, D stand for Identity, CNOT, SWAP and DCNOT unitaries respectively.}
\label{fig2}
\end{figure}

Two-qubit entangling unitaries that can be written as \cite{KC01}
\begin{equation}
U^{AE\to BF}=\sum_{k=1}^{4} e^{-i \lambda_k} \ket{\Lambda_k}\bra{\Lambda_k},
\label{un}
\end{equation}
where $ \ket{\Lambda_k} $ are the so called magic basis states:
\begin{align}
\ket{\Lambda_1}=\frac {1}{\sqrt 2}(\ket{0}_A\ket{0}_E+\ket{1}_A\ket{1}_E), \hspace{10mm}
\ket{\Lambda_2}=\frac {-i}{\sqrt 2}(\ket{0}_A\ket{0}_E-\ket{1}_A\ket{1}_E),
\notag\\
\ket{\Lambda_3}=\frac {1}{\sqrt 2}(\ket{0}_A\ket{1}_E-\ket{1}_A\ket{0}_E),\hspace{10mm}
\ket{\Lambda_4}=\frac {-i}{\sqrt 2}(\ket{0}_A\ket{1}_E+\ket{1}_A\ket{0}_E),
\end{align}
and the eigenvalues $\lambda_k$  are
\begin{align}
\lambda_1=&\frac{\alpha_x-\alpha_y+\alpha_z}{2},   \hspace{12mm} \lambda_2=\frac{-\alpha_x+\alpha_y+\alpha_z}{2},  \notag\\
\lambda_3=&\frac{-\alpha_x-\alpha_y-\alpha_z}{2},  \hspace{9mm}
\lambda_4=\frac{\alpha_x+\alpha_y-\alpha_z}{2},
\end{align}
with
\begin{equation}\label{eq:tetra}
\frac{\pi}{2}\geq \alpha_x \geq \alpha_y \geq \alpha_z \geq 0.
\end{equation}
In the canonical basis $\{\ket{0}_A\ket{0}_E, \ket{0}_A\ket{1}_E, \ket{1}_A\ket{0}_E,\ket{1}_A\ket{1}_E\}$ we have
\begin{equation}\label{Umatrix}
U=\begin{pmatrix}
\cos\left(\frac{\alpha_x-\alpha_y}{2}\right) & 0 & 0 & -i \sin\left(\frac{\alpha_x-\alpha_y}{2}\right) \\
0 & e^{i\alpha_z}\cos\left(\frac{\alpha_x+\alpha_y}{2}\right) & -i e^{i\alpha_z}\sin\left(\frac{\alpha_x+\alpha_y}{2}\right) & 0\\
0 & -i e^{i\alpha_z}\sin\left(\frac{\alpha_x+\alpha_y}{2}\right) & e^{i\alpha_z}\cos\left(\frac{\alpha_x+\alpha_y}{2}\right) & 0\\
-i \sin\left(\frac{\alpha_x-\alpha_y}{2}\right) & 0 & 0 & \cos\left(\frac{\alpha_x-\alpha_y}{2}\right)
\end{pmatrix}.
\end{equation}
All unitaries of this kind form, according to Eq.\eqref{eq:tetra}, a tetrahedron in the parameter space
$\alpha_x,\alpha_y,\alpha_z$ (see Fig. \ref{fig2}). The vertices of such a tetrahedron represent
Identity, SWAP, CNOT and DCNOT unitaries.

%%%%%%%%%%%%%%%%%%%%%%%%%%%%%%%%%%%%%%%%%%%%%%%%%%%%%%%%%%%%%%

\section{One-shot quantum capacity}\label{sec3}

By referring to Fig.\ref{fig1}, for a fixed probe state ${\boldsymbol\phi}$, we can 
consider the channel 
\begin{equation}
 {\cal N}_{\boldsymbol\phi}^{E\to B}({\boldsymbol\theta})={\rm Tr}_F \left( U^{AE\to BF} ({\boldsymbol\phi}\otimes{\boldsymbol\theta}) (U^{AE\to BF})^\dagger  \right),
\end{equation}
and evaluate its capability in transmitting quantum information.
In particular, we are interested to the one-shot quantum capacity for the channel $\left({\cal N}_{\boldsymbol\phi}^{E\to B}({\boldsymbol\theta})\right)^{\otimes n}$.
This will tell us how much quantum information we can extract from system $E^n$ by accessing system $B^n$ in one shot  and with a finite probability of error.

\subsection{Lower bound}

Notice preliminarily that the complementary channel of ${\cal N}_{\boldsymbol\phi}^{E\to B}({\boldsymbol\theta})$ reads
\begin{equation}
 ({\cal N}_{\boldsymbol\phi}^{E\to B})^c({\boldsymbol\theta})={\rm Tr}_B \left( U^{AE\to BF} ({\boldsymbol\phi}\otimes{\boldsymbol\theta}) (U^{AE\to BF})^\dagger  \right).
\end{equation}
Let us also set
\begin{equation}\label{eq:rhoBF}
  \rho_{BF}:=({\rm id}^{A\to B}\otimes ({\cal N}_{\boldsymbol\phi}^{E\to B})^c)(\Phi_{AE}),
\end{equation}
where $\Phi_{AE}$ is maximally entangled state across the systems $A$ and $E$.
Analogously, it will be 
\begin{equation}\label{eq:rhoBnFn}
\rho_{B^nF^n}:= \left({\rm id}^{A^n\to B^n}\otimes \left(({\cal N}_{\boldsymbol\phi}^{E\to B})^c\right)^{\otimes n}\right)(\Phi_{A^nE^n}), 
\end{equation}
with $\Phi_{A^nE^n}$ a maximally entangled state across the systems $A^n$ and $E^n$.

\bigskip

For a given $\epsilon>0$, the one-shot quantum capacity of $\left({\cal N}_{\boldsymbol\phi}^{E\to B}\right)^{\otimes n}$ results as \cite{BD10}
\begin{equation}\label{eq:Qeps}
  Q^{\epsilon}\left(\left({\cal N}_{\boldsymbol\phi}^{E\to B}\right)^{\otimes n}\right)=\max \left\{\log_2 m \vert\, F_{\min}
\left(\left({\cal N}_{\boldsymbol\phi}^{E\to B}\right)^{\otimes n},m\right)\geq 1-\epsilon\right\},
\end{equation}
where
\begin{equation}
 F_{\min}\left(\left({\cal N}_{\boldsymbol\phi}^{E\to B}\right)^{\otimes n}, m\right):=
 \max_{\stackrel{ \mathcal{H}'_E\subset\mathcal{H}^n_E,}{ \dim(\mathcal{H}'_E)=m} }
 \max_{\mathcal{D}} \min_{\vert\omega\rangle\in\mathcal{H}'_E} \langle \omega\vert \left(\mathcal{D}
 \circ \left({\cal N}_{\boldsymbol\phi}^{E\to B}\right)^{\otimes n} \right)(\omega)\vert \omega\rangle.
\end{equation}\\
Here ${\cal D}$ is a decoding map from ${\cal H}_B^n\simeq {\cal H}_E^n$ to ${\cal H}'_B\simeq {\cal H}'_E$.

A lower bound to \eqref{eq:Qeps} is given as follows \cite{PRM18}
\begin{equation}\label{re:oneshot}
  Q^{\epsilon}\left(\left({\cal N}_{\boldsymbol\phi}^{E\to B}\right)^{\otimes n}\right)
  \geq \sup_{\delta\in(0, \sqrt{\epsilon/2})}  \left( H_{\min}^{\sqrt{\epsilon/2}-\delta}\left(B^n\vert F^n\right)_{\rho^n}-4\log_2{\frac{1}{\delta}}-2  \right).
\end{equation}
It is
\begin{equation}
  H_{\min}^\epsilon({B}^n\vert F^n)_{\rho^n}:=\max_{\rho_{B^nF^n}'\in {\cal B}^\epsilon(\rho_{B^nF^n})} 
  H_{\min}({B}^n\vert F^n)_{\rho_{B^nF^n}'},
\end{equation}
where
\begin{equation}
  {\cal B}^{\epsilon}(\rho_{B^nF^n})=\left\{ \rho'_{B^nF^n}\vert \,  {\rm tr}(\rho_{B^nF^n}')\leq 1 \, ,
  \sqrt{1-F(\rho_{B^nF^n}, \rho_{B^nF^n}')}\leq \epsilon
  \right\},
\end{equation}
with
\begin{equation}
  F(\rho_{B^nF^n}, \rho'_{B^nF^n}):=\left\Vert \sqrt{\rho_{B^nF^n}}\sqrt{\rho'_{B^nF^n}}\right\Vert_1^2.
\end{equation}
Furthermore
\begin{equation}\label{eq:Hmin}
  H_{\min}(B^n\vert F^n)_{\rho^n}=\max_{\sigma_{F^n}}\sup\left\{ \lambda\in \mathbb{R}\vert\,  \rho_{B^nF^n}
  \leq 2^{-\lambda}  I_{B^n}\otimes \sigma_{F^n}  \right\},
\end{equation}
with  $\rho_{B^nF^n}$ given by \eqref{eq:rhoBnFn}.

We also have the following inequality for $\alpha\in (1,\infty)$ \cite{T16}
\begin{equation}\label{re:mininequality}
  H_{\min}^\epsilon(B^n\vert F^n)_{\rho^n}\geq \mathbb{H}_q(B^n\vert F^n)_{\rho^n}-\frac{g(\epsilon)}{q-1},
\end{equation}
being
\begin{equation}\label{eq:defg}
g(\epsilon):=-\log_2\left(1-\sqrt{1-\epsilon^2}\right).
\end{equation}
Here $\mathbb{H}_q$ is the conditional Renyi entropy, which is defined as
\begin{equation}\label{re:Renyientropy}
  \mathbb{H}_q(B^n\vert F^n)_{\rho^n}:=\max_{\sigma_{F^n}}-D_q(\rho_{B^nF^n}\Vert I_{B^n}\otimes \sigma_{F^n}),
\end{equation}
with
\begin{equation}\label{re:relative}
  D_q(\rho_{B^nF^n}\Vert I_{B^n}\otimes \sigma_{F^n}):=\frac{1}{q-1} \log_2{\rm Tr}
  \left\{\left[ (I_{B^n}\otimes \sigma_{F^n})^{\frac{1-q}{2q}}
  \rho_{B^nF^n} (I_{B^n}\otimes \sigma_{F^n})^{\frac{1-q}{2q}}  \right]^q    \right\}.
\end{equation}
%%%
Considering  $q=2$, by combining relations \eqref{re:oneshot} and \eqref{re:mininequality}, we have
\begin{equation}
   Q^{\epsilon}\left(\left({\cal N}_{\boldsymbol\phi}^{E\to B}\right)^{\otimes n}\right)
   \geq \sup_{\delta\in(0, \sqrt{\epsilon/2})}  \left[ \mathbb{H}_2({B}^n\vert F^n)_{\rho^n}-g\left(\sqrt{\frac{\epsilon}{2}}-\delta\right)-4\log_2{\frac{1}{\delta}}-2  \right].
\end{equation}
By restricting the attention to product states 
$\sigma_{F^n}=\sigma_F^{\otimes n}$ and $\rho_{B^nF^n}=\rho_{BF}^{\otimes n}$ 
in $\mathbb{H}_2(B^n\vert F^n)_{\rho^{ n}}$ and using $\mathbb{H}_2(B^n\vert F^n)_{\rho^{\otimes n}}\geq n\mathbb{H}_2(B\vert F)_{\rho}$, we finally arrive to
\begin{align}\label{re:asymtoticlower}
  Q^{\epsilon, n}\left({\cal N}_{\boldsymbol\phi}^{E\to B}\right)&:=\frac{1}{n} Q^{\epsilon}\left(\left(
  {\cal N}_{\boldsymbol\phi}^{E\to B}\right)^{\otimes n}\right) \notag \\
& \geq \sup_{\delta\in(0, \sqrt{\epsilon/2})}  \left[ \mathbb{H}_2({B}\vert F)_{\rho}-\frac{1}{n}\left(
g\left(\sqrt{\frac{\epsilon}{2}}-\delta\right)+4\log_2{\frac{1}{\delta}}+2  \right)\right] \notag\\
&=   \mathbb{H}_2({B}\vert F)_{\rho}-\frac{1}{n}\left(
g\left(\sqrt{\frac{\epsilon}{2}}-\delta_\star\right)+4\log_2{\frac{1}{\delta_\star}}+2  \right),
\end{align}
where $\delta_\star$ is the value of $\delta$ minimizing the expression
$g\left(\sqrt{\epsilon/2}-\delta\right)-4\log_2\delta$ within the interval $(0, \sqrt{\epsilon/2})$ (see Appendix \ref{AA}).

%%%%%%%%%%%%%%%%%%%%%%%%%%%%%%%%%%%%%%%%%%%%%%

\subsection{Evaluation on the vertices}\label{sec3b}

We now work out the calculation of the r.h.s. of \eqref{re:asymtoticlower} for the four cases of unitaries in the vertices of tetrahedron (see Fig.\ref{fig2}).

%%%%%%%%%%%%%%%%%%%%%%%%%%%%%%%%%%%%%%%%%%%%%%

\begin{itemize}

\item[i)] $\alpha_x=\alpha_y=\alpha_z=0$ ($U=I$). In this case $\rho_{BF}=\Phi_{AE}$.
Then, we have
   \begin{align}
  \mathbb{H}_2({B}\vert F)_\rho & =-\min_{\sigma_F}\log_2{\rm Tr}\left\{\left( (I_B\otimes \sigma_F)^{-\frac{1}{2}}
  \Phi_{AE}   \right)^2   \right\}\\
    &\leq -\log_2{\rm Tr}\left\{\left(  \Phi_{AE} \right)^2    \right\}\leq 0.
  \end{align}
Therefore the lower bound \eqref{re:asymtoticlower} trivially becomes zero.

%%%%%%%%%%%%%%%%

\item[ii)] $\alpha_x=\frac{\pi}{2}$, $\alpha_y=\alpha_z=0$ ($U={CNOT}$). In this case
 \begin{equation}
  \rho_{BF}=\frac{1}{4}\left(
               \begin{array}{cccc}
                 1 & i\sin(2\phi_1)\cos\phi_2 & \sin(2\phi_1)\sin\phi_2  & 1 \\
                 -i\sin(2\phi_1)\cos\phi_2 & 1 & 1 & -\sin(2\phi_1)\sin\phi_2  \\
                 -\sin(2\phi_1)\sin\phi_2  & 1  & 1 & -\sin(2\phi_1)\cos\phi_2 \\
                 1  & i\sin(2\phi_1)\sin\phi_2 & i\sin(2\phi_1)\cos\phi_2 &  1\\
               \end{array}
             \right).
             \end{equation}
Let us set $\eta_F:=\sqrt{\sigma_F}/{\rm Tr}\sqrt{\sigma_F}$ and assume that $\lambda_1$ and $\lambda_2$ are  eigenvalues of $\sigma_F$ with
 \begin{equation}
   \eta_F= \frac{1}{2}\left(
   \begin{array}{cc}
     1+r_3 & r_1-ir_2 \\
     r_1+ir_2 & 1-r_3\\
   \end{array}\right), \quad
      \sigma_F=\frac{1}{2} \left(
   \begin{array}{cc}
     1+p_3 & p_1-ip_2 \\
     p_1+ip_2 & 1-p_3\\
   \end{array}\right) .
 \end{equation}
   We have
 \begin{align}
  \mathbb{H}_2({B}\vert F)_\rho&=\max_{\sigma_F}\left(-\log_2{\rm Tr}\left\{\left( (I_B\otimes \sigma_F)^{-\frac{1}{2}}\rho_{BF} \right)^2   \right\} \right)\label{l1} \\
    &= -\min_{\sigma_F}\log_2 \frac{{\rm Tr}\left\{   \left( (I_B\otimes \eta_F)^{-1}\rho_{BF}  \right)^2   \right\}}{({\rm Tr}\sqrt{\sigma_F})^2} \label{l2} \\
    &=-\min_{\sigma_F}\log_2  \frac{2(1+r_1^2+(1-r_1^2)\cos^2\phi_1\cos^2\phi_2\sin^2(2\phi_1))}{({\rm Tr}\sqrt{\sigma_F})^2(1-r_1^2-r_2^2-r_3^2)^2}  \label{l3} \\
    &\leq -\min_{\sigma_F}\log_2  \frac{1+r_1^2}{8(({\rm Tr}\sqrt{\sigma_F})^2\det(\eta_F))^2} \label{l4}   \\
    &=  -\min_{\sigma_F}\log_2  \frac{({\rm Tr}\sqrt{\sigma_F})^2(1+r_1^2)}{8\det(\sigma_F)}  \label{l5}  \\
        &  = \max_{\sigma_F}-\log_2  \frac{(\sqrt{\lambda_1}+\sqrt{\lambda_2})^2(1+r_1^2)}{8\lambda_1\lambda_2}
        \label{l6} \\
    &\leq -\min_{\sigma_F}\log_2  \frac{(1+2\sqrt{\lambda_1\lambda_2})(1+r_1^2)}{8\lambda_1\lambda_2}\leq 0.
      \end{align}
For the last inequality we used the fact that $1+2\sqrt{x(1-x)}\geq 8x(1-x)$ for $0\leq x\leq 1$.

Again the lower bound \eqref{re:asymtoticlower} trivially becomes zero.

%%%%%%%%%%%%%%%

\item[iii)] $\alpha_x=\alpha_y=\alpha_z=\frac{\pi}{2}$ ($U={SWAP}$). In this case
\begin{equation}
  \rho_{BF}=\left(
               \begin{array}{cccc}
                 \frac{1}{2}\cos^2(\phi_1) & \frac{1}{4}e^{i\phi_2}\sin(2\phi_1) & 0 &  0\\
                 \frac{1}{4} e^{-i\phi_2}\sin(2\phi_1) & \frac{1}{2}\sin^2(\phi_1) & 0 & 0 \\
                 0 & 0 & \frac{1}{2}\cos^2(\phi_1) & -\frac{1}{4} e^{i\phi_2}\sin(2\phi_1) \\
                 0 & 0 & \frac{1}{4} e^{-i\phi_2}\sin(2\phi_1) & \frac{1}{2}\sin^2(\phi_1)\\
               \end{array}
             \right).
\end{equation}
This can also be written as $\rho_{BF}=\frac{I}{2}\otimes \vert \phi'\rangle\langle\phi'\vert$ where $\vert \phi'\rangle=\cos\phi_1\vert 0\rangle+ e^{i\phi_2}\sin\phi_1\vert 1\rangle$.
Then, we have
\begin{align}
  \mathbb{H}_2(B\vert F)_\rho&=-\min_{\sigma_F}\log_2{\rm Tr}\left\{\left( (I_B\otimes \sigma_F)^{-\frac{1}{4}}
  \left(\frac{I}{2}\otimes \vert \phi'\rangle\langle\phi'\vert\right) (I_B\otimes \sigma_F)^{-\frac{1}{4}}  \right)^2   \right\}\\
    &=-\min_{\sigma_F}\log_2{\rm Tr}\left(\frac{I_B}{4}\right){\rm Tr}\left\{\left(  \sigma_F^{-\frac{1}{4}}\vert \phi'\rangle\langle\phi'\vert \sigma_F^{-\frac{1}{4}}  \right)^2    \right\}\\
    &=1-2\min_{\sigma_F}\log_2 \langle\phi'\vert \sigma_F^{-\frac{1}{2}}\vert\phi'\rangle\geq 0.
  \end{align}
Therefore,   $\max_{\sigma_F} \mathbb{H}_2(B\vert F)_\rho=1$ is attained when $\phi_1=\phi_2=0$ and
$\sigma_F=(1/2)[(1+r_3)\vert0\rangle\langle 0\vert+(1-r_3)\vert 1\rangle\langle 1\vert]$ for $r_3\to 1$.

As a result, the lower bound \eqref{re:asymtoticlower} reads
\begin{equation}
  1-\frac{1}{n}g\left(\sqrt{\frac{\epsilon}{2}}-\delta_\star\right)-\frac{4}{n}\log_2{\frac{1}{\delta_\star}}-\frac{2}{n}.
\end{equation}

%%%%%%%%%%%%%%%%%%%%%%%%%%%%%%%%

\item[iv)] $\alpha_x=\alpha_y=\frac{\pi}{2}$, $\alpha_z=0$ ($U={DCNOT}$). In this case
\begin{equation}
  \rho_{BF}=\left(
               \begin{array}{cccc}
                 \frac{1}{2}\cos^2(\phi_1) & \frac{1}{4}i e^{i\phi_2}\sin(2\phi_1) & 0 &  0\\
                 -\frac{1}{4}i e^{-i\phi_2}\sin(2\phi_1) & \frac{1}{2}\sin^2(\phi_1) & 0 & 0 \\
                 0 & 0 & \frac{1}{2}\cos^2(\phi_1) & -\frac{1}{4}i e^{i\phi_2}\sin(2\phi_1) \\
                 0 & 0 & \frac{1}{4}i e^{-i\phi_2}\sin(2\phi_1) & \frac{1}{2}\sin^2(\phi_1)\\
               \end{array}
             \right).
\end{equation}
This can also be written as $\rho_{BF}=\frac{I}{2}\otimes \vert \phi'\rangle\langle\phi'\vert$ where $\vert \phi'\rangle=\cos\phi_1\vert 0\rangle+ie^{i\phi_2}\sin\phi_1\vert 1\rangle$. Then we can repeat the reasoning of case (iii)
and arrive to  the lower bound \eqref{re:asymtoticlower} as
\begin{equation}
  1-\frac{1}{n}g\left(\sqrt{\frac{\epsilon}{2}}-\delta_\star\right)-\frac{4}{n}\log_2{\frac{1}{\delta_\star}}-\frac{2}{n}.
\end{equation}

\end{itemize}

Summarizing, for I and CNOT no quantum information can be retrieved according to the used figure of merit. Instead for SWAP and DCNOT maximal quantum information retrieval (1 qubit) can be approached by increasing the number of shots $n$ even with a fixed value of error $\epsilon$.

%%%%%%%%%%%%%%%%%%%%%%%%%%%%%%%%%%%%%

\subsection{Evaluation on the edges}

We now extend, with the help of numerics, the analysis of the figure of merit along the edges of the tetrahedron.

\begin{figure}[H]
\begin{center}
          \includegraphics[width=0.65\textwidth]{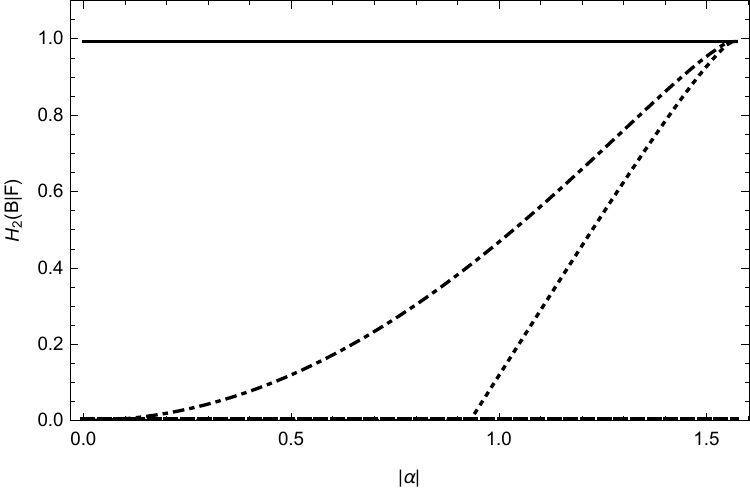}
          \caption{Quantity $\mathbb{H}_2(B\vert F)_\rho$ evaluated along the various edges of the tetrahedron as function of the parameter $|\boldsymbol{\alpha}|$. Dashed curve corresponds to the edge IC; dotted curve to the edges IS and ID;  dashed-dotted curve to the edges CD and CS; solid curve to the edge DS.}
\label{fig3}
\end{center}
\end{figure}

In the cases analyzed in SubSec.\ref{sec3b} the relevant part is given by $\max_{\sigma_F}\mathbb{H}_2(B\vert F)_\rho$. 
The continuity of $\mathbb{H}_2(B\vert F)_\rho$ (see Appendix \ref{AC}), hence its uniform continuity,
allows us to reliable sampling discrete points for plotting its behavior along edges of the tetrahedron of Fig.\ref{fig2}.
The results are summarized in Fig.\ref{fig3}.

It is worth noticing that, while along the edge IC the quantity $\max_{\sigma_F}\mathbb{H}_2(B\vert F)_\rho$  
remains always zero, along the edges IS, ID it becomes nonzero only after a certain (threshold) value of $|\boldsymbol{\alpha}|$ ($|\boldsymbol{\alpha}|\approx\pi/3.4)$.
Instead, along the edges CD, CS it increases with a smooth derivative since the CNOT.  Lastly, for the edge DS it is constant and equal to 1.

The results of this SubSection indicate that there should be a nonzero volume of unitaries around identity for which quantum information retrieval would not be possible.

%%%%%%%%%%%%%%%%%%%%%%%%%%%%%%%%%%%%%%%%%%%%%%%%%%%%%%%%%%%%%%

\section{Quantum state estimation}\label{sec4}

By referring to Fig.\ref{fig1}, we now consider the goal of 
estimating parameters $\boldsymbol{\theta}=(r,\theta_1,\theta_2)\in\Theta$, where
$\Theta=[0\leq r \leq \frac{1}{2}, 0\leq\theta_1\leq\pi, 0\leq \theta_2 \leq 2\pi]$, characterizing the environment state
\begin{equation}
\boldsymbol{\theta}=\frac{1}{2}I+r \sin{\theta_1}\cos{\theta_2}\sigma_x+r \sin{\theta_1}\sin{\theta_2}\sigma_y+r \cos{\theta_2}\sigma_z.
\end{equation}
Here $\sigma_x,\sigma_y,\sigma_z$ are the Pauli matrices. 
The estimation should be done
by means of the channel output $\cN_{\boldsymbol{\theta}} (\boldsymbol{\phi})$.
As a figure of merit we will derive a Bayesian version of the quantum Cramer-Rao bound following
the classical bound \cite{GL95}.

\subsection{Lower bound}

Let $\{ p\left(x;\boldsymbol{\theta}\right) \}_{\boldsymbol{\theta}\in\Theta} $ be a family of probability density functions on a sample space $\cal X$. Any unbiased estimator $\hat{\boldsymbol{\theta}}$ of $\boldsymbol{\theta}$ satisfies the multivariate Cramer-Rao inequality:
\be
\mathbb{E}_{p}\left[\left(\hat{\bb\theta}-\bb\theta\right)\left(\hat{\bb\theta}-\bb\theta\right)^T\right]\geq
{\cal I}(\bb\theta)^{-1},
\label{CR}
\ee
where $\mathbb{E}_{p}$ denotes the expectation with respect to $ p\left(x;\bb\theta\right)$, ${\cal I}(\bb\theta)$ is the classical Fisher information, and
$\mathbb{E}_{p}\left[\left(\hat{\bb\theta}-\bb\theta\right)\left(\hat{\bb\theta}-\bb\theta\right)^T\right]$
is the covariance matrix of $\hat{\bb\theta}$.

After assigning a proper prior probability distribution $\pi$ to $\bb\theta$ and taking the expectation of Eq.\eqref{CR}, we can get the simplest Bayesian Cramer-Rao bound:
\be
\mathbb{E}_{\pi}\left\{
\mathbb{E}_{p}\left[\left(\hat{\bb\theta}-\bb\theta\right)\left(\hat{\bb\theta}-\bb\theta\right)^T\right]\right\}\geq
\mathbb{E}_\pi\left\{\left[{\cal I}\left({\bb\theta}\right)\right]\right\}^{-1}.\label{BCR}
\ee

We assume the parameters $r$, $\theta_1$ and $\theta_2$ independent, and distributed according to the measure \cite{ZS01} $\pi({\boldsymbol\theta})d{\boldsymbol\theta}=\sin\frac{\theta_1}{2} {\rm d} r {\rm d}\theta_1{\rm d}\theta_2/(2\pi)$. Taking into account that quantum Fisher information ${\cal{F}}({\bb\theta})$ upper bounds the classical Fisher information ${\cal{I}}({\bb\theta})$, we get
\bea
\mathbb{E}_{\pi}\left\{
\mathbb{E}_{p} \left[\left(\hat{\bb\theta}-{\bb\theta}\right)\left(\hat{\bb\theta}-{\bb\theta}\right)^T\right]\right\}&\geq& \left\{\mathbb{E}_\pi\left[{\cal I}({\bb\theta})\right]\right\}^{-1} \notag\\
&=& \frac{4}{\int_{\Theta}{\rm{tr}}\{{\cal{I}}({\bb\theta})\}{\pi}({\bb\theta}){\rm{d}}{\bb\theta}}\notag\\
&\geq& \frac{4}{\int_{\Theta}{\rm{tr}}\{{\cal{F}}({\bb\theta})\}{\pi}({\bb\theta}){\rm{d}}{\bb\theta}}.
\label{VTree}
\eea

Then the quantity of interest for us becomes
\begin{equation}\label{F}
F_{\bb\phi}:=\int_{\Theta}{\rm{tr}}\{{\cal{F}}({\bb\theta})\}{\pi}({\bb\theta}){\rm{d}}{\bb\theta},
\end{equation}
where the subscript emphasizes its dependence from the input state $\bb\phi$ of the $A$ system.

Its maximum overall states $\bb\phi$ will be denoted by $\overline{F}$.  Due to the convexity of the Fisher information,  the optimization can be restricted to pure states of the form
\be
\ket{\boldsymbol{\phi}}=\cos\frac{\phi_1}{2}\ket 0+e^{i\phi_2}\sin\frac{\phi_1}{2}\ket 1,
\ee
where
$\phi_1\in[0,\pi]$ and $ \phi_2 \in[0, 2\pi]$.

%%%%

From Ref.\cite{Liu20}, we know that the basis-independent expression of quantum Fisher information for a single-qubit mixed state $\rho({\bb\theta})$ reads
\be\label{Fab}
{\cal{F}}_{ab}={\rm tr}[\left( \partial_{a} \rho\right)\left( \partial_{b} \rho\right)]
+\frac{{\rm tr}[\rho \left(\partial_{a} \rho\right)\rho\left( \partial_{b} \rho\right)]}{{\rm det}\left(\rho\right)},
\ee
where $\partial_a\rho=\partial\rho/\partial\theta_a$.
For a pure qubit state Eq.\eqref{Fab} reduces to
\be
{\cal{F}}_{ab}=2{\rm tr}[\left( \partial_{a} \rho\right)\left( \partial_{b} \rho\right)].
\ee

%%%%%%%%%%%%%%%%%%%%%%%%%%%%%%%%%%%%%%%

\subsection{Evaluation on the vertices}

We now work out the calculation of $\overline{F}$ for the four cases of unitaries in the vertices of the tetrahedron (see Fig.\ref{fig2}).

\begin{itemize}

\item[i)] $\alpha_x=\alpha_y=\alpha_z=0$ ($U=I$). In this case Eq.\eqref{Umatrix} reduces to the identity and hence ${\cal F}=0$ for all input $|{\boldsymbol\phi}\rangle$. As a consequence $\overline{F}=0$.

%%%

\item[ii)] $\alpha_x=\frac{\pi}{2}$, $\alpha_y=\alpha_z=0$ ($U=CNOT$). In this case 
\begin{align}
{\cal N}_{\boldsymbol\theta}({\boldsymbol{\phi}})&=\frac{1}{2} +r\sin\theta_1\cos\theta_2\sin\phi_1\sin\phi_2
\ket 0 \bra 0\notag\\
&+\frac{1}{2}\sin\phi_1\cos\phi_2+i r\sin\theta_1\cos\theta_2\cos\phi_1\ket 0\bra 1 \notag\\
&+\frac{1}{2}\sin\phi_1\cos\phi_2-i r\sin\theta_1\cos\theta_2\cos\phi_1 \ket1 \bra 0\notag\\
&+\frac{1}{2}-r\sin\theta_1\cos\theta_2\sin\phi_1\sin\phi_2
|1\rangle\langle 1|.
\end{align}
Then
\bea
  {\cal F}_{\theta_1,\theta_1}&=&\frac{4 r^2 \cos ^2\theta_1\cos ^2\theta_2 \left(1- \sin ^2\phi_1 \cos^ 2 \phi_2\right)}{1-4 r^2\sin^2\theta_1\cos^2\theta_2},\notag\\
  {\cal F}_{\theta_2,\theta_2}&=&\frac{4 r^2 \sin ^2\theta_1\sin ^2\theta_2 \left(1- \sin ^2\phi_1 \cos^ 2 \phi_2\right)}{1-4 r^2\sin^2\theta_1\cos^2\theta_2},\notag\\
   {\cal F}_{r,r}&=&\frac{ 4 \sin ^2\theta_1\cos ^2\theta_2 \left(1- \sin ^2\phi_1 \cos^ 2 \phi_2\right)}{1-4 r^2\sin^2\theta_1\cos^2\theta_2}.
\eea
The maximum of
\be
F_{\bb\phi}=\int_{\Theta}{\rm{Tr}}\{{\cal{F}}({\bb\theta})\}{\pi}({\bb\theta}){\rm{d}}{\bb\theta}.
\ee
is achieved for $\phi_1=0,\pi$ or $\phi_2=\frac{\pi}{2},\frac{3\pi}{2}$.
In this case it is  $\overline{F}=1.76108$.

%%%%

\item[iii)] $\alpha_x=\alpha_y=\alpha_z=\frac{\pi}{2}$ ($U=SWAP$). In this case 
\bea
{\cal N}_{\boldsymbol\theta}({\boldsymbol{\phi}})&=&\left(\frac{1}{2}+r \cos\theta_1\right)\ket 0\bra 0
+r e^{-i\theta_2}\sin\theta_1  |0\rangle\langle 1|\notag\\
&+& r e^{i\theta_2}\sin\theta_1 |1\rangle\langle 0|
+\left(\frac{1}{2}-r \cos\theta_1\right)\ket 1 \bra 1.
\eea
Then
\bea
&&{\cal F}_{\theta_1,\theta_1}=4 r^2 ,\\
&&{\cal F}_{\theta_2,\theta_2}=4r^2\sin^2\theta_1,\\
&&{\cal F}_{r,r}=\frac{4}{1-4 r^2}.
 \eea
Clearly
\be
\int_{\Theta}{\rm{Tr}}\{{\cal{F}}({\bb\theta})\}{\pi}({\bb\theta}){\rm{d}}{\bb\theta},
\ee
diverges, meaning that for any state $\bb\phi$ we will have a good estimation on average.

%%%%

\item[iv)] $\alpha_x=\alpha_y=\frac{\pi}{2}$, $\alpha_z=0$ ($U=DCNOT$). In this case 
\begin{align}
{\cal N}_{\boldsymbol\theta}({\boldsymbol{\phi}})&=\left(\frac{1}{2}+r \cos\theta_1\right) \ket 0 \bra 0
+ i e^{-i \theta_2} r \sin \theta_1 \cos \phi_1 \ket 0\bra 1\notag \\
&-i e^{i \theta_2} r \sin \theta_1 \cos \phi_1 \ket 1\bra 0
+\left(\frac{1}{2}-r \cos\theta_1\right)\ket 1 \bra 1.
\end{align}
Then
\bea
{\cal F}_{\theta_1,\theta_1}&=&\frac{4r^2\left(\sin^2 \theta_1-(4 r^2-\cos^2\theta_1)\cos^2\phi_1\right)}{1-4 r^2 \left( \cos^2\theta_1+\sin^2{\theta_1}\cos^2{\phi_1}\right)}, \\
{\cal F}_{\theta_2,\theta_2}&=& 4 r^2\sin^2{\theta_1}\cos^2{\phi_1}, \\
{\cal F}_{r,r}&=&\frac{4\left( \cos^2\theta_1+\sin^2{\theta_1}\cos^2{\phi_1}\right)}{1-4 r^2 \left( \cos^2\theta_1+\sin^2{\theta_1}\cos^2{\phi_1}\right)}.
\eea
Notice that these three terms are non negative, hence maximizing \eqref{F} is equivalent to maximize them.
However, their derivative with respect to $\cos^2\phi_1$ never nullify.
Hence, the maximum is at the extreme points. Actually it is easy tho see that it occurs for $\cos^2\phi_1=1$.

As a consequence we will have
\begin{equation}
\overline{F}=\int_{\Theta} \left[4r^2+4r^2\sin^2\theta_1+ \frac{4}{1-4r^2}\right] {\pi}({\bb\theta}){\rm{d}}{\bb\theta},
\end{equation}
which diverges, likewise the case of SWAP.

\end{itemize}

\bigskip

The fact that $\overline{F}$ diverges for DCNOT and SWAP is in line with the results in Sec.\ref{sec3}
where the bound for one-shot capacity in these cases turns out to be (close to) 1.
In other words for DCNOT and SWAP the environment state can be estimated with perfect accuracy 
or analogously it can be transmitted with maximum reliability to the $B$ system.

Also for identity we have concordance between the two approaches. In fact the environment state can be estimated with total inaccuracy 
or analogously it can be transmitted to the $B$ system in a total unreliable way.

The situation is slightly different for CNOT, since according to the one-shot capacity approach it behaves like identity,
while state environment estimation can be done with a finite average error.   

%%%%%%%%%%%%%%%%%%%%%%%%%%%%%%%%%%%%%%%

\subsection{Evaluation on the edges}

We now extend, with the help of numerics, the analysis of the figure of merit along the edges of the tetrahedron.

\begin{figure}[H]
\begin{center}
          \includegraphics[width=0.65\textwidth]{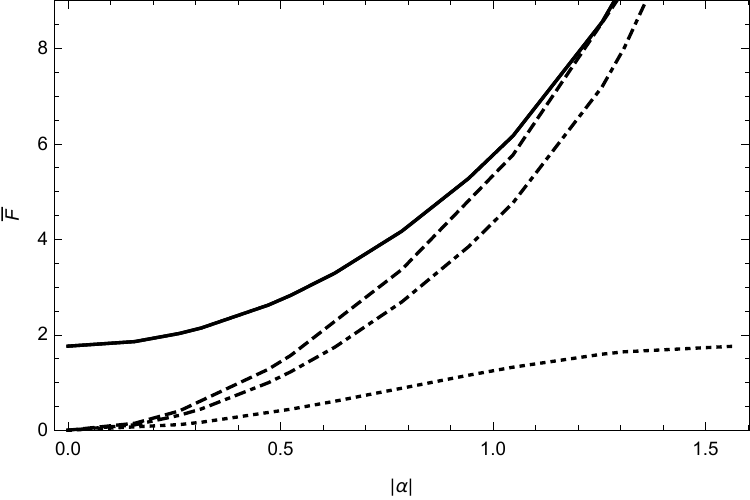}
          \caption{Quantity $\overline{F}$ evaluated along the various edges of the tetrahedron as function of the parameter 
          $|\boldsymbol{\alpha}|$. Curves from bottom to top refer respectively to the edges
          (IC), (ID), (IS), (CS and CD). The edge (DS) is not represented because along it, 
          the value of $\overline{F}$ is infinity.}
\label{fig4}
\end{center}
\end{figure}

The continuity of the average quantum Fisher information (see Appendix \ref{AB}), hence its uniform continuity,
allows us to reliable sampling discrete points for plotting its behavior along edges of the tetrahedron of Fig.\ref{fig2}.
The results are summarized in Fig.\ref{fig4}.

From Fig.\ref{fig4} it is left out the edge (SD). On that edge, we have \eqref{Umatrix} as
\begin{equation}\label{edge}
U=\begin{pmatrix}
  e^{-\frac{i z}{2} }& 0 & 0 & 0 \\
0 & 0 & -i e^{\frac{i z}{2}} & 0 \\
0 &-i e^{\frac{i z}{2}} & 0 & 0 \\
0 & 0 & 0 & e^{-\frac{i z}{2}}
\end{pmatrix}
\end{equation}
Therefore
\bea
{\cal N}_{\boldsymbol\theta}({\boldsymbol{\phi}})&=&\left(\frac{1}{2}+r\cos \theta_1\right)\ket 0\bra 0+r e^{-i \theta_2}\sin{\theta_1} (\sin\alpha_ z+i \cos\alpha_ z \cos \phi_1)\ket 0\bra 1\notag\\
&+&
r e^{i \theta_2}\sin{\theta_1} (\sin\alpha_ z-i \cos\alpha_ z \cos (\phi_1))\ket 1\bra 0+\left(\frac{1}{2}-r\cos \theta_1\right)\ket 1\bra 1, \notag\\
\eea
and
\bea
{\cal F}_{\theta_1,\theta_1}&=&\frac{4r^2\left(1-4r^2+(4 r^2-\cos^2\theta_1)\sin^2\phi_1\cos^2\alpha_z\right)}{1-4 r^2 \left(1- \cos^2\alpha_z\sin^2{\theta_1}\sin^2{\phi_1}\right)},\\
{\cal F}_{\theta_2,\theta_2}&=& 4 r^2\sin^2{\theta_1}(1-\cos^2{\alpha_ z}\sin^2{\phi_1}),\\
{\cal F}_{r,r}&=&\frac{4\left(1- \cos^2\alpha_ z\sin^2{\theta_1}\sin^2{\phi_1}\right)}{1-4 r^2 \left( 1-\cos^2\alpha_ z\sin^2{\theta_1}\sin^2{\phi_1}\right)}.
\eea
$F_{{\boldsymbol{\phi}}}$ does not have extrema with respect to $\alpha_z$ inside the interval $\left(0,\frac{\pi}{2}\right)$, and we know that on the two ends of the interval it has the same maximum value. Thus, we can conclude, that on the edge SD
we have the same value of $\overline{F}$, i.e. infinity.

\bigskip

The results of this SubSection indicate that only for identity quantum information retrieval would not be possible. For all other unitaries it would be possible, although in some cases with a finite average error.

%%%%%%%%%%%%%%%%%%%%%%%%%%%%%%%%%%%%%%%%%%%%%%%%%%%%%%%%%%%%%

\section{Conclusion}

In conclusion, we have addressed the problem of reading quantum information by a quantum probe, thus going beyond the standard paradigm that confines quantum reading to the retrieval of classical information. As a model of quantum memory we used environment parametrized quantum channels arising from two-qubit unitaries.
Since these unitaries lie in a tetrahedron in $\mathbb{R}^3$, we characterized those on the edges to have general insights. To this end, we used a lower bound to the one-shot quantum capacity of the channel connecting the environment with the output of the main system (probe) as well as a Bayesian version of the quantum Cramer-Rao bound for the initial environment state.
We remark that while the first also showed the behavior vs the number $n$ of shots, the second
just refers to one-shot,
Notwithstanding, the results of the first are more restrictive.
In fact, according to the first figure of merit, there should be a nonzero volume of unitaries around identity for which quantum information retrieval would not be possible. Instead, the second shows that only for identity quantum information retrieval would not be possible. For all other unitaries it would be possible, although in some cases with a finite average error. 

This difference of behavior between the two figures of merit (to most striking of which occurs for CNOT) should be ascribed to the non tightness of the used lower bound for to the one-shot quantum capacity.

In future, besides investigating also the unitaries inside the tetrahedron, we plan to characterize spatial arrays of the presented model to figure out the performance of realistic quantum memories. A fact that can be useful for applications in quantum cryptography as well as in quantum computation.

%%%%%%%%%%%%%%%%%%%%%%%%%%%%%%%%%%%%%%%%%%%%%%%%%%%%%%%%%%%%%%

\appendix

%%%%%%%%%%%%%%%%%%%%%%%%%%%%%%%%%%%%%%%%%%

\section{Value of $\delta_\star$}\label{AA}

\begin{align}
\delta_\star&=\frac{13 \sqrt{\epsilon}}{15 \sqrt{2}} \notag\\
&-\frac{\left(1+i \sqrt{3}\right) \sqrt[3]{\sqrt{2 \epsilon^3}+\sqrt{1728 \epsilon^2+715392 \epsilon-5971968}
+648 \sqrt{2 \epsilon }}}{30\ 2^{2/3}}\notag\\
&-\frac{\left(1-i \sqrt{3}\right) (\epsilon+144)}{30 \sqrt[3]{2} \sqrt[3]{\sqrt{2\epsilon^3}+\sqrt{1728 \epsilon^2+715392 \epsilon-5971968}+648 \sqrt{2\epsilon } }}.
\end{align}

%%%%%%%%%%%%%%%%%%%%%%%%%%%%%%%%%%%%%%%%%%%%%%

\section{Continuity of lower bound on the one-shot quantum capacity}\label{AC}

Let us consider
\begin{align}
  \rho_{BF}&=\left({\rm id}_{A\to B} \otimes \left({\cal N}_{E\to B}\right)^c\right)(\Phi_{AE}), \\
    \rho'_{BF}&=\left({\rm id}_{A\to B} \otimes \left({\cal N}'_{E\to B}\right)^c\right)(\Phi_{AE}),
\end{align}
where ${\cal N}$ and ${\cal N}^\prime$ refer respectively to the unitaries $U_{\boldsymbol{\alpha}}$ and 
$U_{\boldsymbol{\alpha}'}$. Then, we have 
\begin{align}
 \left\|  \rho_{BF}  - \rho'_{BF} \right\|_1 & = \left\|\left({\rm id}_{A\to B} \otimes \left(\left({\cal N}_{E\to B}\right)^c -\left({\cal N}'_{E\to B}\right)^c \right) \right)(\Phi_{AE}) \right\|_1\\
 & \leq   \left\|\left({\cal N}_{E\to B}\right)^c - \left({\cal N}'_{E\to B}\right)^c \right\|_\diamond \\
& \leq C_1 \left\|  U_{\boldsymbol{\alpha}} - U_{\boldsymbol{\alpha}'}\right\|_\infty \label{re:74}\\
&\leq C_2  \left\| {\boldsymbol{\alpha}}-{\boldsymbol{\alpha}'} \right\|_{\mathbb{R}^3},
\end{align}
where $C_i<+\infty$ are positive constants. Furthermore, $\| \cdot \|_\diamond$ is the diamond norm 
and relation \eqref{re:74} comes from \cite{KSW08}.  

Now, for a given $\epsilon=\left\| {\boldsymbol{\alpha}}-{\boldsymbol{\alpha}'} \right\|_{\mathbb{R}^3}>0$, it is shown \cite{MD22} that
\begin{equation}
 \left\vert \mathbb{H}_2({B}\vert F)_\rho -\mathbb{H}_2({B}\vert F)_{\rho'}\right\vert\leq \log_2(1+\sqrt{\epsilon})+3\log_2\left(1+64\sqrt[3]{\epsilon} -\frac{\sqrt{\epsilon}}{\sqrt[3]{1+\sqrt{\epsilon}}}\right)  ,   \end{equation}
where $\mathbb{H}_2({B}\vert F)_\rho$ is defined in \eqref{re:Renyientropy}.

%%%%%%%%%%%%%%%%%%%%%%%%%%%%%%%%%%%%%%%%%%%%%%%%%%%%%%%%%%%%%

\section{Continuity of average quantum Fisher information}\label{AB}

Let us denote by $\boldsymbol{\alpha}$ the parameters vector characterizing the unitary.

\begin{align}
\left| \overline{F}_{\boldsymbol{\alpha}}-\overline{F}_{\boldsymbol{\alpha}'}\right| &=\left|
\max_{\boldsymbol{\phi}}\int_\Theta {\rm Tr}\{{\cal F}_{\boldsymbol{\alpha}}({\boldsymbol{\theta}},{\boldsymbol{\phi}})\} \pi({\boldsymbol{\theta}})
d{\boldsymbol{\theta}}
-\max_{\boldsymbol{\phi}}\int_\Theta {\rm Tr}\{{\cal F}_{\boldsymbol{\alpha}'}({\boldsymbol{\theta}},{\boldsymbol{\phi}})\} \pi({\boldsymbol{\theta}})
d{\boldsymbol{\theta}}\right|
\label{Fcontinuity1} \\
&\leq \max_{\boldsymbol{\phi,\phi'}}
\left|
\int_\Theta {\rm Tr}\{{\cal F}_{\boldsymbol{\alpha}}({\boldsymbol{\theta}},{\boldsymbol{\phi}})\} \pi({\boldsymbol{\theta}})
d{\boldsymbol{\theta}}
-\int_\Theta {\rm Tr}\{{\cal F}_{\boldsymbol{\alpha}'}({\boldsymbol{\theta}},{\boldsymbol{\phi'}})\} \pi({\boldsymbol{\theta}})
d{\boldsymbol{\theta}}\right|
\label{Fcontinuity2} \\
& \leq \max_{\boldsymbol{\phi,\phi'}}
\int_\Theta \left| {\rm Tr}\{{\cal F}_{\boldsymbol{\alpha}}({\boldsymbol{\theta}},{\boldsymbol{\phi}})\}
- {\rm Tr}\{{\cal F}_{\boldsymbol{\alpha}'}({\boldsymbol{\theta}},{\boldsymbol{\phi'}})\} \right| \pi({\boldsymbol{\theta}})
d{\boldsymbol{\theta}}
\label{Fcontinuity3} \\
& \leq C_1 \max_{\boldsymbol{\theta,\phi,\phi'}}
\left| {\rm Tr}\{{\cal F}_{\boldsymbol{\alpha}}({\boldsymbol{\theta}},{\boldsymbol{\phi}})\}
- {\rm Tr}\{{\cal F}_{\boldsymbol{\alpha}'}({\boldsymbol{\theta}},{\boldsymbol{\phi'}})\} \right|
\label{Fcontinuity4} \\
& \leq C_2 \max_{\boldsymbol{\theta,\theta',\phi,\phi'}}
\left| {\rm Tr}\{{\cal F}_{\boldsymbol{\alpha}}({\boldsymbol{\theta}},{\boldsymbol{\phi}})\}
- {\rm Tr}\{{\cal F}_{\boldsymbol{\alpha}'}({\boldsymbol{\theta'}},{\boldsymbol{\phi'}})\} \right|
\label{Fcontinuity5} \\
& \leq C_3 \max_{\boldsymbol{\theta,\theta',\phi,\phi'}}
\left\| {\cal N}^{\boldsymbol{\alpha}}_{\boldsymbol{\theta}}({\boldsymbol{\phi}})
- {\cal N}^{\boldsymbol{\alpha}'}_{\boldsymbol{\theta'}}({\boldsymbol{\phi'}}) \right\|_1
\label{Fcontinuity6} \\
& = C_3 \max_{\boldsymbol{\theta,\theta',\phi,\phi'}}
\left\| {\rm Tr}_E \left[ U_{\boldsymbol{\alpha}} {\boldsymbol{\phi}}\otimes {\boldsymbol{\theta}} U_{\boldsymbol{\alpha}}^\dag\right]
- {\rm Tr}_E \left[ U_{\boldsymbol{\alpha'}} {\boldsymbol{\phi'}}\otimes {\boldsymbol{\theta'}} U_{\boldsymbol{\alpha'}}^\dag\right]
\right\|_1
\label{Fcontinuity7} \\
& \leq C_4 \max_{\boldsymbol{\theta,\theta',\phi,\phi'}}
\left\| U_{\boldsymbol{\alpha}} {\boldsymbol{\phi}}\otimes {\boldsymbol{\theta}} U_{\boldsymbol{\alpha}}^\dag
- U_{\boldsymbol{\alpha'}} {\boldsymbol{\phi'}}\otimes {\boldsymbol{\theta'}} U_{\boldsymbol{\alpha'}}^\dag
\right\|_1
\label{Fcontinuity8} \\
& \leq C_5 \left\| U_{\boldsymbol{\alpha}}-U_{\boldsymbol{\alpha'}} \right\|_{\infty}\\
&\leq C_6  \left\| {\boldsymbol{\alpha}}-{\boldsymbol{\alpha}'} \right\|_{\mathbb{R}^3}
\label{Fcontinuity9}
\end{align}
where $C_i <+\infty$ are positive constants.
In going from \eqref{Fcontinuity5} to \eqref{Fcontinuity6} we used the continuity of the Fisher information \cite{RA19}.
In going from \eqref{Fcontinuity7} to \eqref{Fcontinuity8} we used the the property that discarding a system cannot increase the norm \cite{Rast12}.

%%%%%%%%%%%%%%%%%%%%%%%%%%%%%%%%%%%%%%%%%%%%%%%%%%%%%%%%%%%%%%

\end{document}